\documentclass[aip,jap,reprint,groupaddress,noshowpacs]{revtex4-1}
\pdfoutput=1

\usepackage[colorlinks=true,bookmarks=true]{hyper ref}

\usepackage{mathrsfs}
\usepackage{epsfig}
\usepackage{amsmath}
\usepackage{subfigure}
\usepackage{amsfonts}

\begin{document}

\title{A novel workflow for seismic net pay estimation with uncertainty}
\author{Michael E. Glinsky}
\author{Dale Baptiste}
\author{Muhlis Unaldi}
\affiliation{Geotrace Technologies, Houston, TX, USA}
\author{Vishal Nagassar}
\affiliation{Centrica E\&P, Port of Spain, Trinidad}

\begin{abstract}
This paper presents a novel workflow for seismic net pay estimation with uncertainty.  It is demonstrated on the Cassra/Iris Field.  The theory for the stochastic wavelet derivation (which estimates the seismic noise level along with the wavelet, time-to-depth mapping, and their uncertainties), the stochastic sparse spike inversion, and the net pay estimation (using secant areas) along with its uncertainty; will be outlined.  This includes benchmarking of this methodology on a synthetic model.  A critical part of this process is the calibration of the secant areas.  This is done in a two step process.  First, a preliminary calibration is done with the stochastic reflection response modeling using rock physics relationships derived from the well logs.  Second, a refinement is made to the calibration to account for the encountered net pay at the wells.  Finally, a variogram structure is estimated from the extracted secant area map, then used to build in the lateral correlation to the ensemble of net pay maps while matching the well results to within the nugget of the variogram.  These net pay maps are then integrated, over the area of full saturation gas, to give the GIIP distribution (Gaussian distributions for the porosity, gas expansion factor, and gas saturation for the sand end member are assumed and incorporated in the estimate of GIIP).  The method is demonstrated on the Iris (UP5 turbidite) interval.  The net pay is corrected for reduction in the amplitudes over part of the area due to shallow gas.  The sensitivity of the GIIP to the independent stochastic variables is estimated (determining the value of information) so that business decisions can be made that maximize the value of the field.
\end{abstract}

\maketitle

\section{Introduction}

This paper presents the initial implementation and demonstration of a novel work flow.  Certain parts are established technology using previously developed, but not widely used methods (i.e., stochastic wavelet derivation, sparse spike inversion based on the ideas of Daubechies and Mallat, net pay estimation using secant area, stochastic reflection response modeling using rock physics relationships derived from well logs, and geostatistical simulation of lateral correlation taking into account well measurements).  This not withstanding, significant development was needed to assimilate these capabilities in an integrated workflow.  Other parts were novel technology that was developed and applied for the first time (i.e., the stochastic aspect of the sparse spike inversion and the net pay estimation).  

The details and synthetic benchmarking of this technology suite will be discussed in Sec. \ref{theory}.  More specifically, the wavelet derivation and seismic noise estimation will be discussed in Sec. \ref{wavelet.derivation}, the novel stochastic sparse spike inversion will be discussed in Sec. \ref{stochastic.ssi}, the secant amplitude extraction and stochastic net pay estimation (that is, net pay with uncertainty) will be discussed in Sec. \ref{net.pay.estimation}, the two step calibration process that includes the stochastic reflection response modeling will be discussed in Sec. \ref{calibration}, and how the lateral correlation and well measurements of net pay are built into the ensemble of net pay estimates will be discussed in Sec. \ref{lateral.correlation}.

\begin{figure}
\noindent\includegraphics[width=20pc]{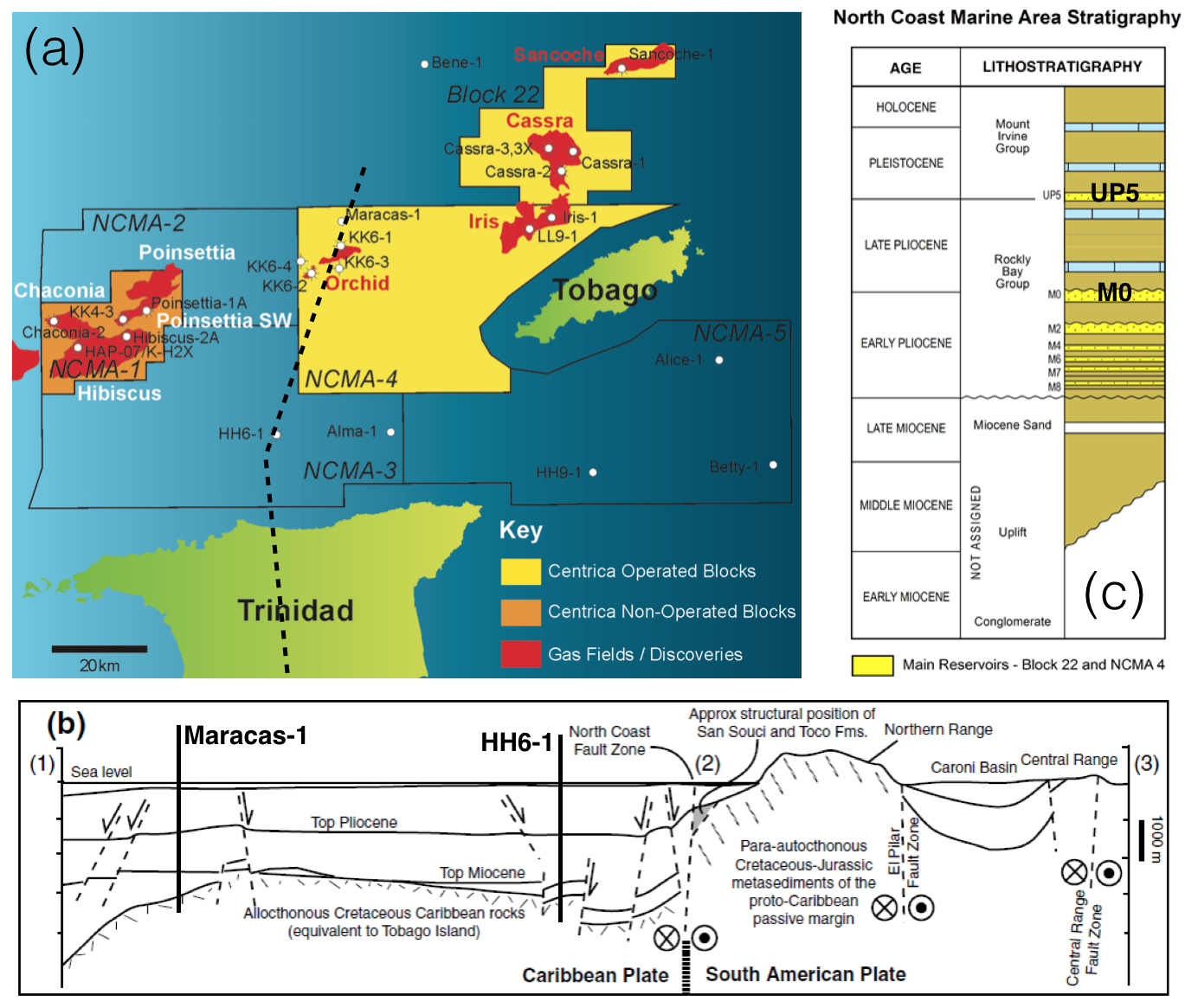}
\caption{\label{geology.fig} The geological context of the Cassra/Iris Field:  (a) map showing location of the field, the dotted line shows the location of the cross section shown in (b) geologic cross section of the field, (c) stratigraphic section indicating the two main reservoir units.}
\end{figure}

This methodology will be demonstrated on the Cassra/Iris Field.  These licenses (Block 22 and NCMA-4) lie on the regional Patao High basement structure located within the West Tobago Basin, offshore the north coast of Trinidad and to the northwest coast of Tobago (see Fig. \ref{geology.fig}). The area contains primarily upper Miocene to Pleistocene aged clastic sediments resting on a heterogeneous basement of Jurassic/Cretaceous age consisting of metamorphic and igneous rocks.

In the Block 22 license, the Cassra gas discovery is contained in the uppermost early Pliocene M0 reservoir sands. In NCMA-4, the Iris gas discovery is in Pleistocene aged reservoir sands, termed UP5. The hydrocarbon system consists of mainly combination structural-stratigraphic traps, formed by compactional drape over basement structural highs, which are sealed by intra-formational shales/silts that also provide the dry biogenic gas source for all the gas discoveries and producing fields of the basin.
%===============================%
%===============================%

Cassra is the main discovered resource in Block 22 with the resource area being clearly defined on 3D seismic data as a strong amplitude anomaly covering around $60 \, \text{km}^2$. The trap is a combination structural-stratigraphic trap in good quality reservoir sands of early Pliocene age, known locally as the M0 reservoir.   Strong clinoformal geometries are seen on seismic and attest to the progradation of smaller-scale ``parasquence-set'' within the overall sand body unit.

Iris straddles the boundary between licenses NCMA-4 and Block 22 and the Iris field is clearly defined on 3D seismic data as a strong amplitude anomaly, covering around $100 \, \text{km}^2$. The trap is a combination structural-stratigraphic trap in Pleistocene aged sands, locally termed UP5, which are interpreted as a deep water channel-lobe complex.

Section \ref{results} presents the results of the analysis on the turbidite sequences of the UP5/Iris interval.  The volumetric distribution is corrected for a reduction in amplitude over part of the area due to shallow gas.

\section{Theory and methodology}
\label{theory}

The theory and the methodology is presented in the following subsections:

\subsection{Wavelet derivation and noise estimation}
\label{wavelet.derivation}

A wavelet derivation engine, originally developed by \citet{gunning.glinsky.06}, was used to estimate the seismic wavelet.  More graphic details can be found in a presentation by \citet{glinsky.06}.  This wavelet derivation derives one wavelet that matches multiple, possibly deviated wells.  It also estimates the vertical time-to-depth mapping, the lateral placement of the seismic, the length of the wavelet, and the seismic noise level.  All of these have their uncertainty estimated.  The results of the application of this will be shown in Sec. \ref{results}.  The wavelet and the seismic noise level are explicitly used to do the stochastic sparse spike inversion described in Sec. \ref{stochastic.ssi}.  The time-to-depth mapping and lateral placement can be used to locate the well on the seismic and to understand the uncertainty in that placement.

\subsection{Stochastic Sparse Spike Inversion (3SI)}
\label{stochastic.ssi}

The basic sparse spike inversion engine behind the Stochastic Sparse Spike Inversion (3SI), was developed by \citet{dossal.mallat.05} based on the concepts of \citet{daubechies.et.al.04}.  It has guaranteed convergence to norm, that is to say that it has no problems with being trapped in local minimums.  This is a very important property of the inversion.  Other industry inversions must build low frequency models, then use these to constrain their results away from the local minimums.  In the process, they consume much project time in building the low frequency model, and bias the result away from the correct solution in the process.  Examples of these destructive behaviors will be shown later in Sec. \ref{net.pay.estimation}.

The results are integrated (by a running sum or runsum, for short), the running mean is removed \citep{glinsky.et.al.09} to eliminate the divergence at small scale.  This result can be directly compared to the logarithm of the acoustic impedance.

The free parameters of this inversion (seismic noise level or spikiness, and the length of the zero mean gate) are estimated by minimizing the deviation of the sparse spike inversion result on the seismic data at the wells from the well log measurements of acoustic impedance.

The uncertainty in this inversion is then estimated by a bootstrap procedure.  An ensemble of inversions are generated by the following recipe.  To generate one member of the ensemble, a wavelet is chosen at random from the ensemble of wavelets generated by the wavelet derivation in Sec. \ref{wavelet.derivation}.  A perturbed version is created of the original seismic by adding band limited white noise to the measured seismic.  The amount of the noise is consistent with the noise level determined by the stochastic wavelet derivation.  The bandwidth of the noise is that of the measured seismic data.  A sparse spike inversion is then done using this version of the wavelet and the seismic data.  As many members of the ensemble, as needed to obtain sufficient statistical accuracy in the distributions, are generated by repeating these steps.  The result of this 3SI can be found in Sec. \ref{results}.

\subsection{Secant amplitude extraction and net pay estimation}
\label{net.pay.estimation}

\begin{figure}
\noindent\includegraphics[width=20pc]{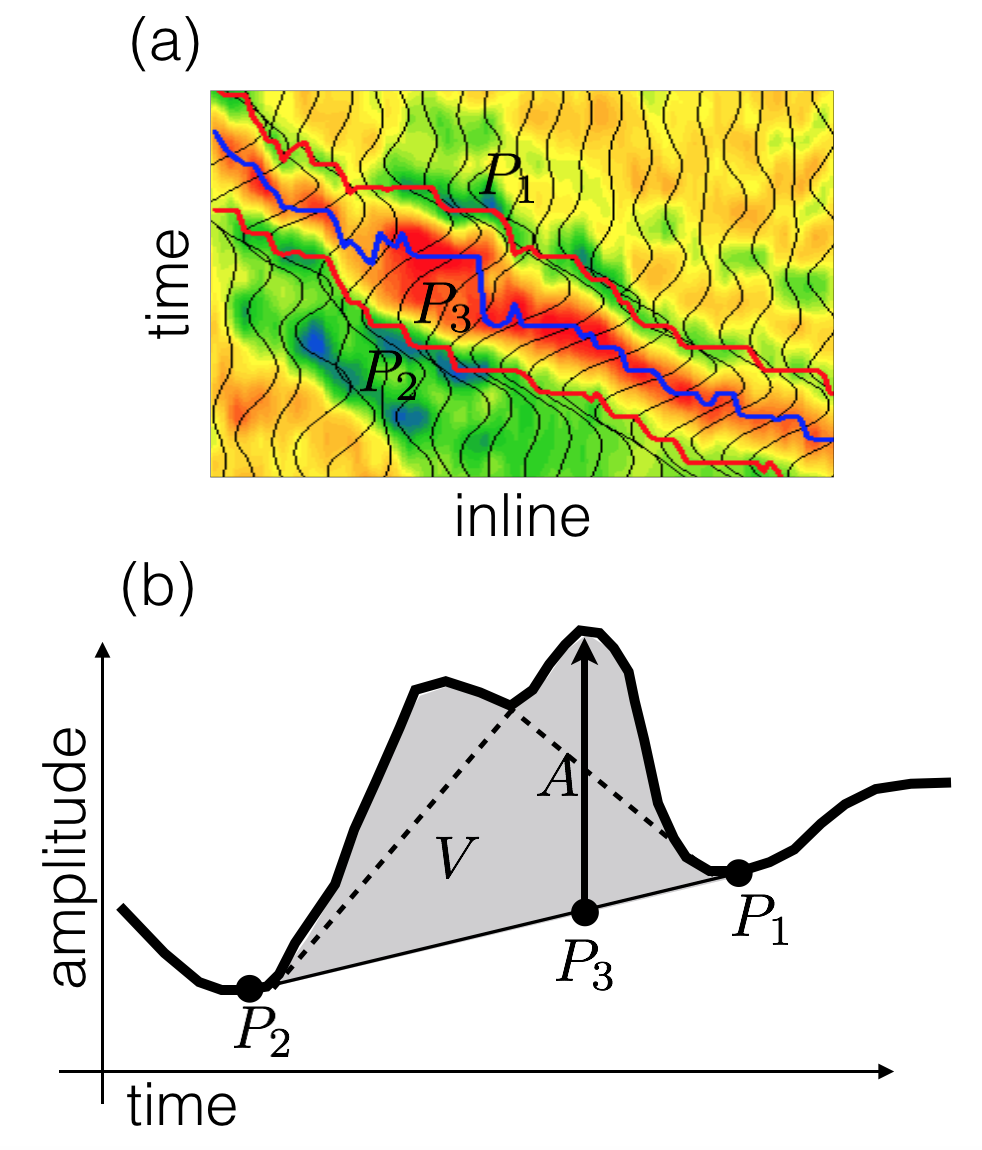}
\caption{\label{secant.fig} Secant amplitude extraction:  (a) location of the three secant points shown as horizons on a cross section of acoustic impedance from a sparse spike inversion.  Shown is the original seismic data as the black wiggles.  The false color image is the acoustic impedance of the sparse spike inversion using a rainbow colorbar where soft impedances are the hot red colors and the hard acoustic impedances are the cool blue colors, (b) a seismic trace showing the secant line, improper dashed secant lines that attempt to separate the two sand units, the location of the three secant points, the secant amplitude $A$, and the grey secant area $V$.}
\end{figure}

The proper way to extract amplitudes and amplitude thicknesses from ``runsum'' data, is the secant method.  The corresponding amplitude is called the secant amplitude, and the corresponding amplitude thickness is called the secant area.  As shown in Fig. \ref{secant.fig}, the secant line is drawn and the secant points, $P_1$ and $P_2$ are determined.  The maximum height above this secant line is determined at the point $P_3$.  This is the secant amplitude, $A$.  The area above the secant line and below the curve is determined.  This is the secant area, $V$.  The amount of net sand, $S$ is proportional to the secant area, that is
\begin{equation}
S \sim V.
\end{equation}
The constant of proportionality can be expressed two different ways giving either
\begin{equation}
\label{secant.a.eqn}
S = V (N_g / A_0)
\end{equation}
or
\begin{equation}
\label{secant.b.eqn}
S = V (v_s / 2 R_s),
\end{equation}
where $N_g$ is the average net-to-gross of the sand, $A_0$ is the average secant amplitude for regions of the sand above tuning, $v_s$ is the average velocity of the sand, and $R_s$ is the average modeled reflection coefficient of the end member sand.  Eq. \eqref{secant.a.eqn} is best used when there are no well penetrations of the objective, that is a rank exploration case.  An initial estimate in the appraisal case can be estimated by Eq. \eqref{secant.b.eqn}, where $v_s$ and $R_s$ are both estimated from the stochastic reflection modeling of the rock physics relationships.  The appraisal calibration procedure will be discussed in more detail in Sec. \ref{calibration}.  A very good, unbiased estimate of net sand will be shown.  
%===============================%
%===============================%

It should be noted in Fig. \ref{secant.fig} that the secant area is calculated over the total package.  If small sands that are interfering with each other have their amplitudes and areas extracted separately, there will be a significant error made as demonstrated by the dashed lines in the figure.

An advantage in using secant area is that it is insensitive to phase, up to first order in the error in the phase.  It is also an intensive property, not an extensive property like a root mean square (RMS) amplitude.  In other words, the amplitude is not dependent on the length or choice of the interval.  The secant points are uniquely defined and can be determined as part of the algorithm, where the the end points for the RMS are determined outside the algorithm by the analyst.  This makes the RMS amplitude dependent both on the choice of gate and the width of the sand.

\begin{figure}
\noindent\includegraphics[width=20pc]{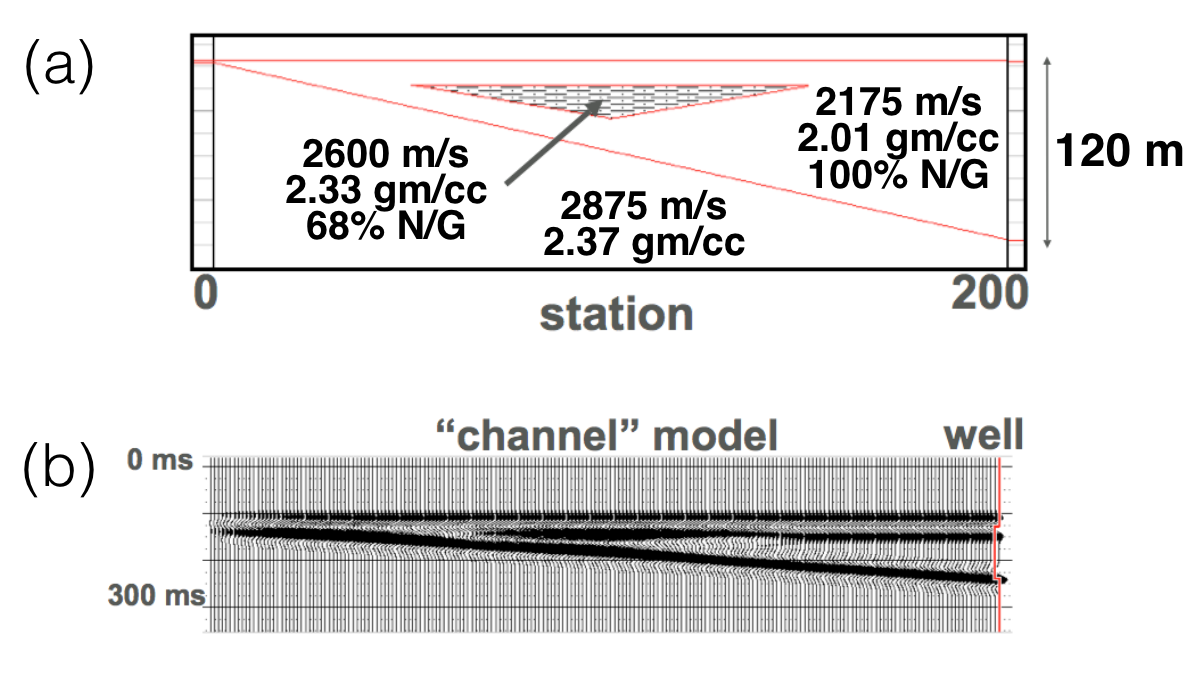}
\caption{\label{wedge.model.fig} Synthetic model used to benchmark the sparse spike inversion, secant area extraction and the net sand calculation: (a) the structure of the model, (b) the synthetic seismic section.}
\end{figure}

\begin{figure}
\noindent\includegraphics[width=20pc]{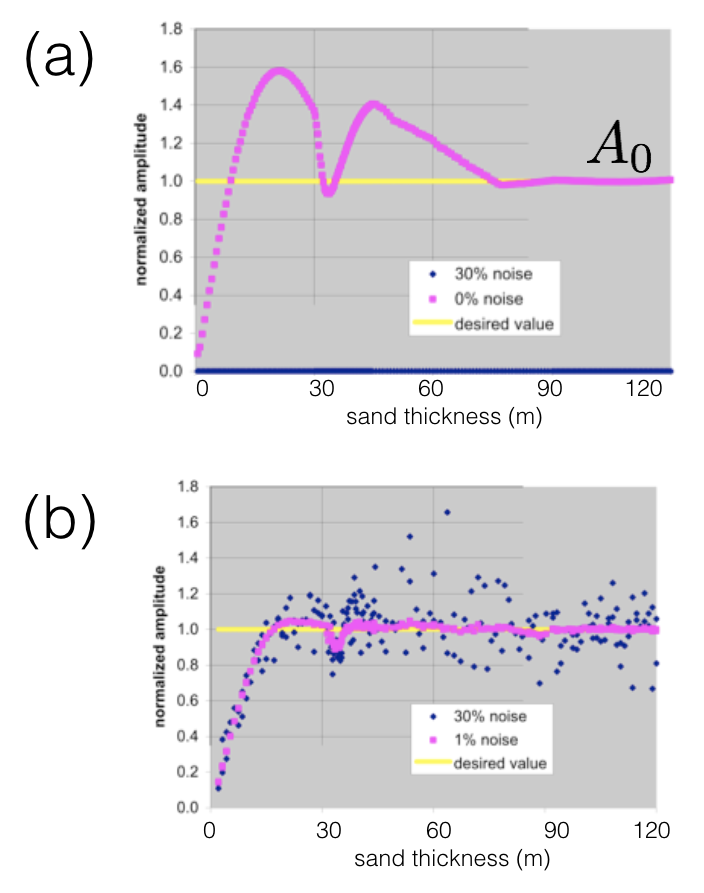}
\caption{\label{secant.amp.fig} Secant amplitude extraction of the synthetic model shown in Fig. \ref{wedge.model.fig}:  (a) extraction from the un-inverted runsum data.  The correct amplitude is shown as the yellow line and $A_0$, (b) extraction from the sparse spike inverted data.}
\end{figure}

\begin{figure}
\noindent\includegraphics[width=20pc]{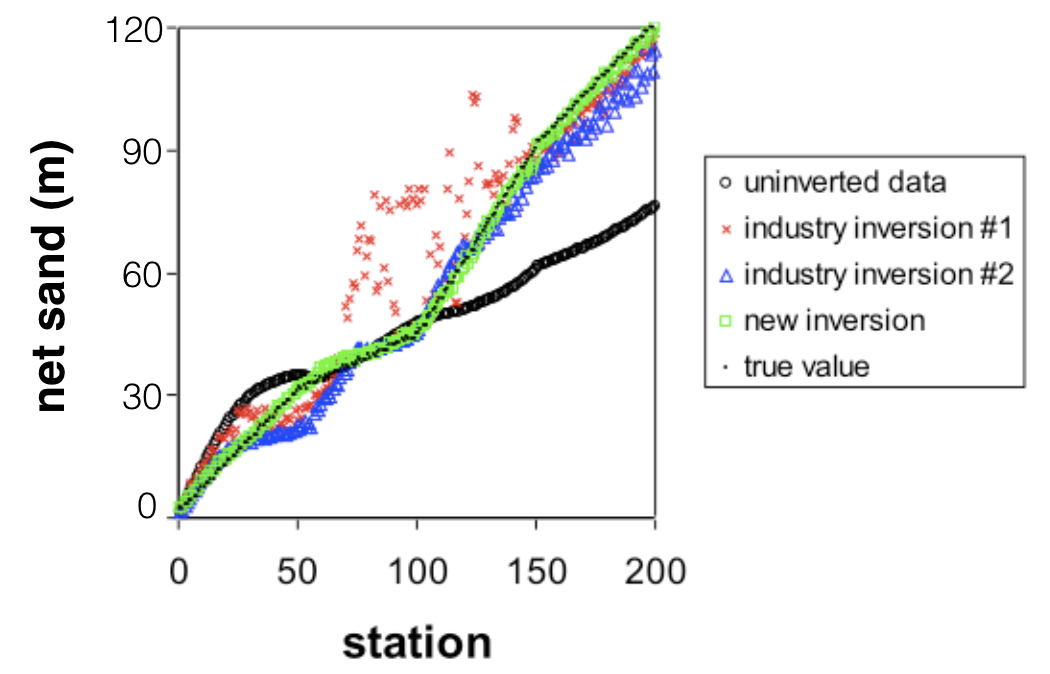}
\caption{\label{net.sand.fig} Net sand estimated from the secant area extraction of the synthetic model shown in Fig. \ref{wedge.model.fig}. Shown are the results for the original seismic runsum data, the true value, the new inversion used by this analysis, and two common industry sparse spike inversions.}
\end{figure}

To benchmark this procedure and the various different types of sparse spike inversions, the model shown in Fig. \ref{wedge.model.fig} was constructed.  It is a soft, type III sand, characteristic of the U.S. Gulf of Mexico plio-pleistocene, imbedded in a uniform shale with a shale lens.  The maximum thickness of this shale lens, and the thickness of the sand on either side of this lens at the maximum lens thickness was set to be exactly the tuning thickness of the 20 Hz Richter wavelet used to generate the synthetic seismic.  This was done to present the greatest challenge to the sparse spike inversions.  The result of the secant amplitude extraction on the ``runsum'' data before and after the sparse spike inversion that we have used are shown in Fig. \ref{secant.amp.fig}.  Note the tuning enhancement before the inversion, and how the inversion has removed that resonance.  The $A_0$ was estimated from this figure, then the secant area extraction along with Eq. \eqref{secant.a.eqn} was used to calculate the net sand.  The result is shown in Fig. \ref{net.sand.fig}.  This was done for the original seismic data, our sparse spike inversion, and two common industry inversions.  Notice the dramatic improvement from the initial seismic data to our sparse spike inversion.  The match of our sparse spike inversion to the true answer is nearly perfect.  The same can not be said for the other two industry inversions.  Both suffer from local minimums and the bias introduced by the low frequency model used to avoid them.
%===============================%
%===============================%
%===============================%
%===============================%
%===============================%
%===============================%

\subsection{Calibration}
\label{calibration}

The calibration was done in two steps.  The preliminary calibration consisted of stochastic reflection modeling given the rock physics relationships.  The average sand velocity, $v_s$, and the average reflection coefficient for the end member sand, $R_s$, were calculated then substituted into Eq. \eqref{secant.b.eqn} to generate the net sand map. 

\begin{figure}
\noindent\includegraphics[width=20pc]{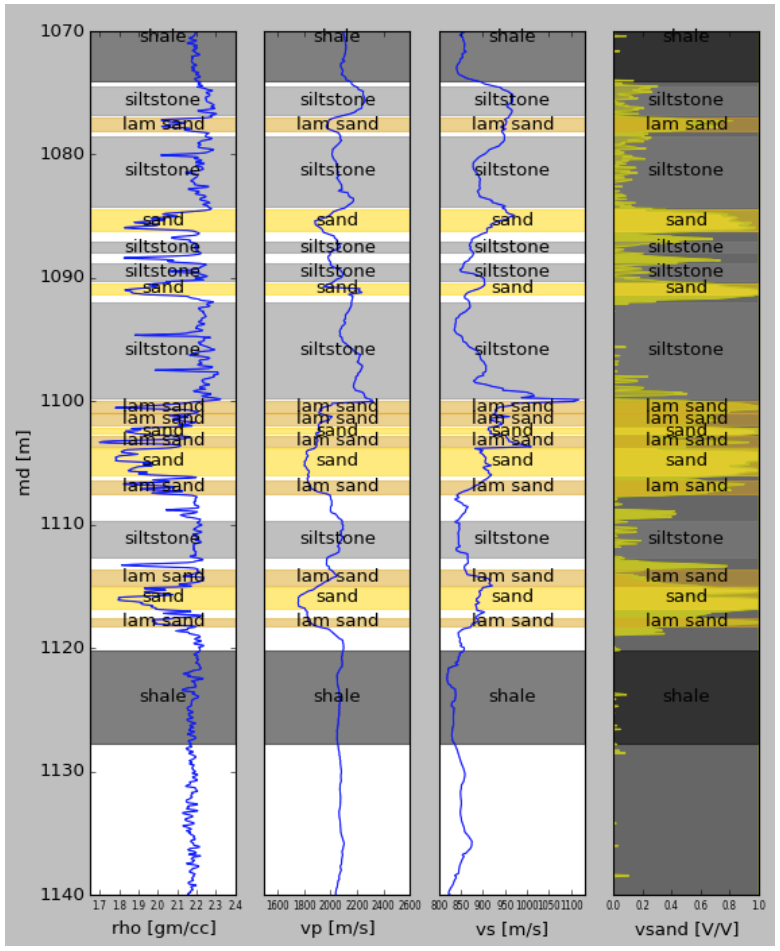}
\caption{\label{rock.phys.petro.fig} Hand picked intervals of end members (sand, laminated sand, shale, and silt) for Iris-1.}
\end{figure}

The well logs were first analyzed to determine the rock physics relationships.  For this analysis a set of intervals were picked by hand that are characteristic of the end-member lithologies, as shown in Fig. \ref{rock.phys.petro.fig}.  A set of linear relationships of the form
\begin{equation}
\label{vp.eqn}
v_p = A_{vp} + B_{vp} \:  \text{TVDBML} + C_{vp} \: \text{LFIV} + \sigma_{vp},
\end{equation}
\begin{equation}
\label{vs.eqn}
v_s = A_{vs} + B_{vs} v_p + \sigma_{vs}, 
\end{equation}
and
\begin{equation}
\label{rho.eqn}
\rho = A_\rho + B_\rho v_p + \sigma_\rho.
\end{equation}
were fit to the acoustic properties, where TVDBML is Total Vertical Depth Below Mud Line, LFIV is Low Frequency Interval Velocity (practically will be seismic imaging velocity, but is a 250 m average of the sonic velocity in this regression), and the $\sigma$'s are the uncertainties in the fits.  Equation \eqref{rho.eqn} is often called the Gardner-Gardner-Gregory relationship \citep{gardner.et.al.74}, and Eq. \eqref{vs.eqn} the Castagna-Greenburg relationship \citep{castagna.et.al.98}.

\begin{figure}
\noindent\includegraphics[width=20pc]{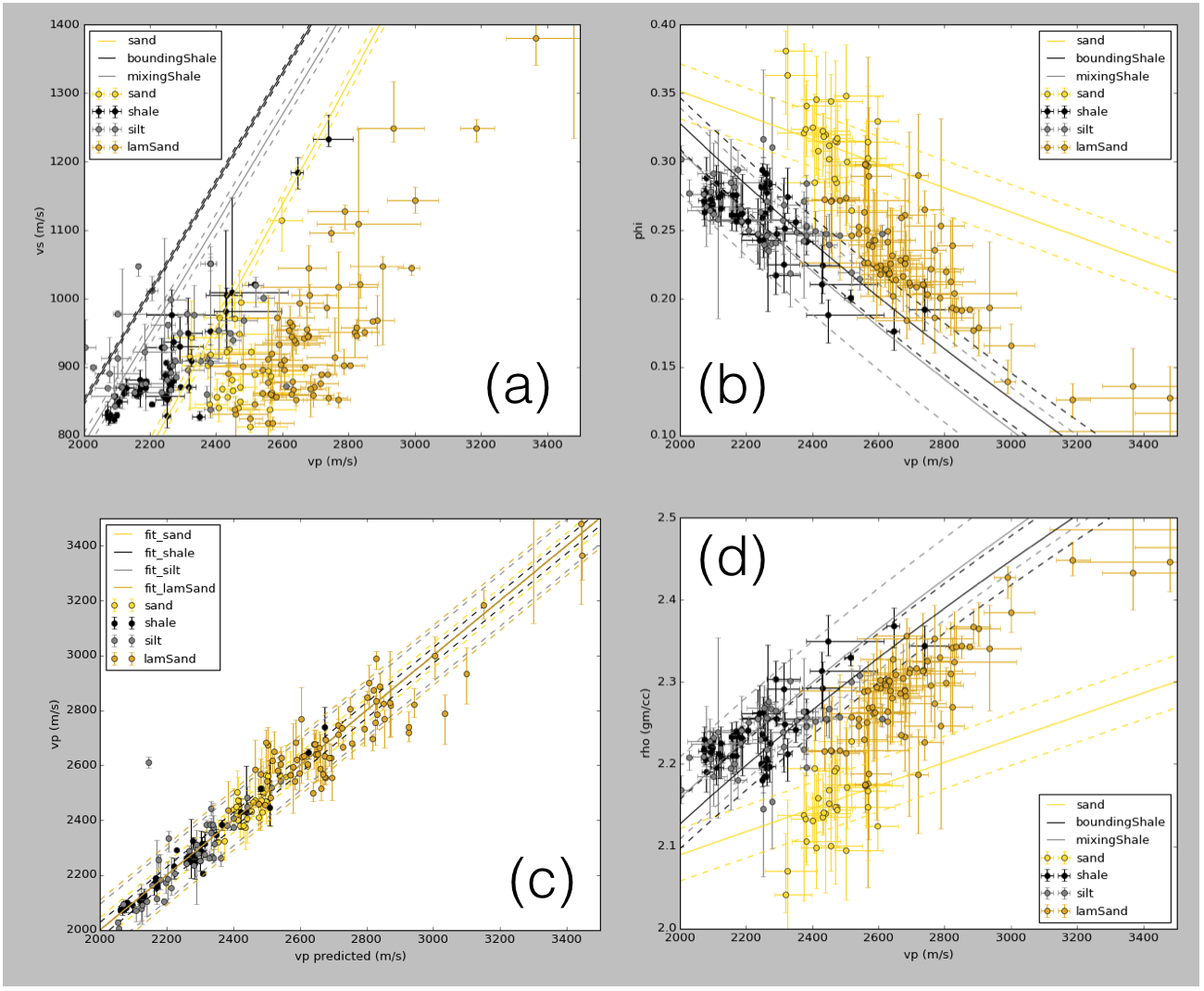}
\caption{\label{rock.phys.fit.fig} Cross plots of the acoustic properties of the end member lithologies including the ``fit'' lines with uncertainty (dotted lines show one standard deviation) used in the subsequent stochastic rock physics modeling:  (a) $v_p$ versus $v_s$ relationship of Eq. \eqref{vs.eqn}, also called the Castagna-Greenburg relationship, (b) porosity versus $v_p$ relationship equivalent to Eq. \eqref{rho.eqn}, (c) predictability of the compaction relationship of Eq. \eqref{vp.eqn}, (d) the density versus $v_p$ relationship of Eq. \eqref{rho.eqn}, also known as the Gardner-Gardner-Gregory relationship.  Points with error bars (one sigma) are taken from five wells (Iris-1, Cassra-1, Cassra-2, Cassra-3, and Cassra-3X).}
\end{figure}

These fits are shown in Fig. \ref{rock.phys.fit.fig}.  Note that end-member sands were fluid substituted to have a reference brine in their pore space.  The fit lines shown in these figures have been modified to be characteristic of only the best sand points and the most characteristic shale points. The slopes have also been modified to be consistent with regional trends.
%===============================%
%===============================%
%===============================%
%===============================%

\begin{figure}
\noindent\includegraphics[width=20pc]{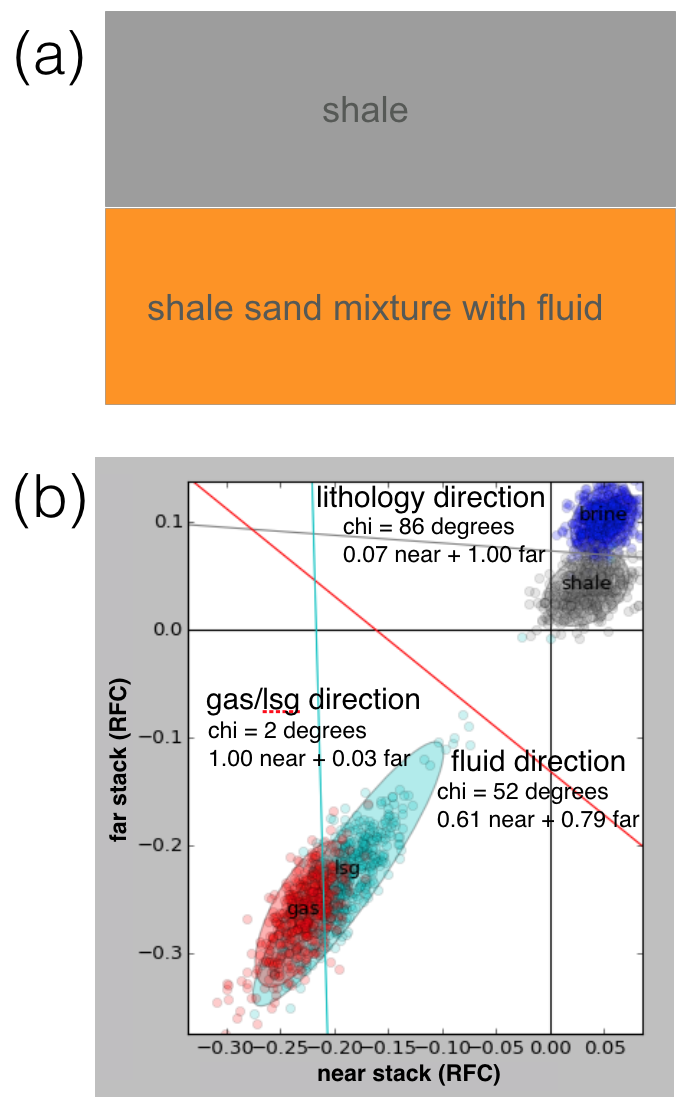}
\caption{\label{stoch.rfc.fig} Stochastic reflection modeling for UP5/Iris:  (a) two-layer model that is modeled, (b) results of the stochastic modeling.  Shown are the distribution functions of the near and far reflection coefficients for three different fluids in the pore space of the sand, and a shale on shale reflector.  An ensemble of points is shown along with a transparent ellipse showing the two sigma range of a fit 2D Gaussian distribution. The optimum lines for separation of two of the clusters are also shown (lithology=shale/brine, fluid=brine/gas, and gas/lsg).}
\end{figure}

An ensemble (with 5000 realizations) of two layer models were then constructed and the reflection response calculated, see Fig. \ref{stoch.rfc.fig}.  The model consisted of an end-member shale overlying a laminated sand.  The laminated sand consisted of the end-member shale mixed with the end-member reference sand with the target fluid in its pore space.
%===============================%
%===============================%

\begin{figure}
\noindent\includegraphics[width=20pc]{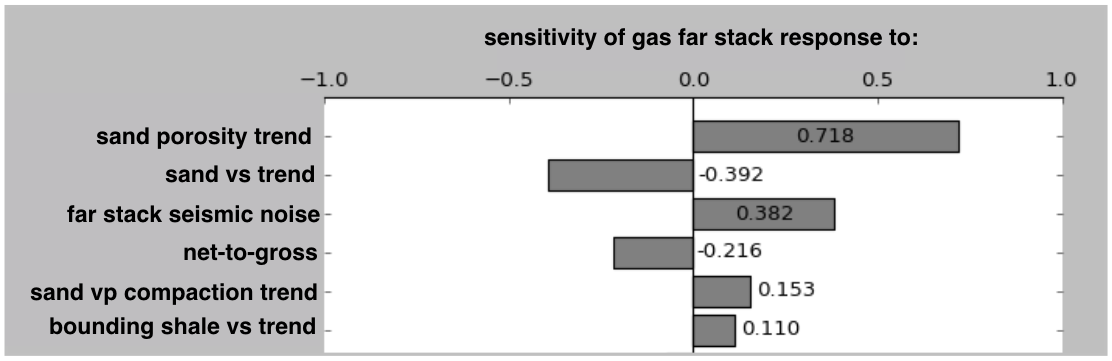}
\caption{\label{rock.phys.sensitivity.fig} Sensitivity of reflection coefficient to the independent variables for UP5/Iris.}
\end{figure}

Also generated by this stochastic reflection modeling is the sensitivity of the reflection response to the input rock physics relationships and other rock parameters (e.g., hydrocarbon saturation, fluid properties, and net-to-gross).  An example of this analysis is shown in Fig. \ref{rock.phys.sensitivity.fig} which shows the tornado chart of Spearman rank order correlation.  This can be very useful information to inform decisions on future technical work and data acquisition.  For the sensitivity shown in Fig. \ref{rock.phys.sensitivity.fig}, the sand porosity trend is at the top of the tornado chart, followed by the sand $v_s$ trend, the far stack seismic noise, and finally the net-to-gross.  If a smaller uncertainty in the volumetric estimation is needed to make important development decisions more petrophysical work should be done on analysis of the density logs, followed by better shear log acquisition, then work on the seismic processing to improve the signal-to-noise level, and finally geologic study to improve the estimate of the net-to-gross.
%===============================%
%===============================%

The distribution of the expected reflection responses generated by this procedure, can be used as conditional probabilities in a Bayesian estimation of fluid probabilities given an observed reflection response of secant amplitudes.  The result is a set of fluid probability maps.  It should be noted that the sand needs to be of tuning thickness or greater.  Interpretation of these probabilities should recognize that sand below tuning or with a net-to-gross less than what was modeled will have secant amplitudes less than what would be modeled for that fluid.

The performance of this classifier is quantified by calculating the confusion matrix shown in Table \ref{avo.confusion.table}.    The confusion matrix is the matrix of probabilities of classification, $P(\text{class}_i | \text{class}_j)$, of $\text{class}_i$ given that it is $\text{class}_j$.  This matrix shows that there is some confusion in separating full saturation from low saturation gas.  Gas is only correctly identified 70\% of the time.  All the other cases are well separated and correctly identified 96\% to 100\% of the time.  The value in multiple stack AVO analysis is quantified by examining the confusion matrix for only the full stack (see Table \ref{confusion.table}) and comparing it to confusion matrix for using the near and the far stack previously calculated in Table \ref{avo.confusion.table}.  While gas is still easily distinguished from either shale or brine, AVO does lead to better discrimination of brine from shale.
%===============================%
\begin{table}
\centering
\begin{tabular}{|c||c|c|c|c|}
\hline
$\text{class}_j  \setminus \text{class}_i$ & \; $\mathbf{shale}$ \; & \; $\mathbf{brine}$ \;  & \; $\mathbf{lsg}$ \; & \; $\mathbf{gas}$ \; \\
\hline \hline
$\mathbf{shale}$ & 96 & 4 & 0 & 0\\
\hline
$\mathbf{brine}$ & 4 & 96 & 0 & 0\\
\hline
$\mathbf{lsg}$ & 0 & 2 & 62 & 36\\
\hline
$\mathbf{gas}$ & 0 & 0 & 30 & 70\\
\hline
\end{tabular}
\caption{Confusion matrix for UP5/Iris,  $P(\text{class}_i | \text{class}_j)$, probability of classification as $\text{class}_i$ given that it is $\text{class}_j$, using both the near and far stacks (i.e., multiple stack AVO analysis).  Values are in percent.}
\label{avo.confusion.table}
\end{table}
%===============================%
%===============================%
\begin{table}
\centering
\begin{tabular}{|c||c|c|c|c|}
\hline
$\text{class}_j   \setminus \text{class}_i$ & \; $\mathbf{shale}$ \; & \; $\mathbf{brine}$ \;  & \; $\mathbf{lsg}$ \; & \; $\mathbf{gas}$ \; \\
\hline \hline
$\mathbf{shale}$ & 59 & 41 & 0 & 0\\
\hline
$\mathbf{brine}$ & 41 & 59 & 0 & 0\\
\hline
$\mathbf{lsg}$ & 1 & 1 & 61 & 37\\
\hline
$\mathbf{gas}$ & 0 & 0 & 32 & 68\\
\hline
\end{tabular}
\caption{Confusion matrix for UP5/Iris,  $P(\text{class}_i | \text{class}_j)$, probability of classification as $\text{class}_i$ given that it is $\text{class}_j$, using only the full stack.  Values are in percent.}
\label{confusion.table}
\end{table}
%===============================%

\begin{figure}
\noindent\includegraphics[width=20pc]{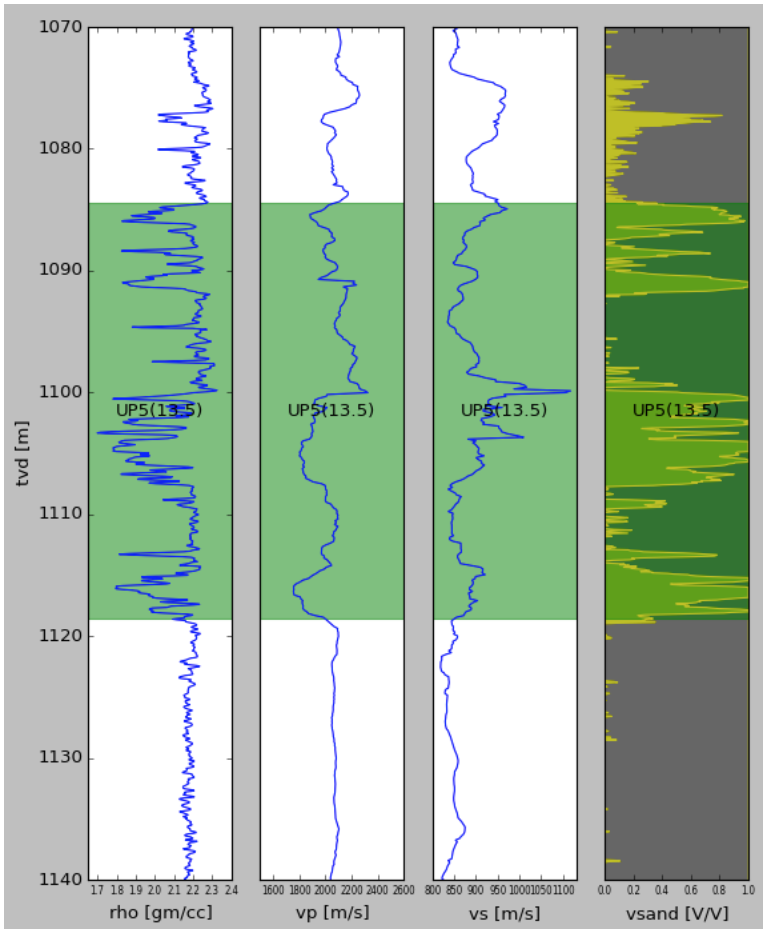}
\caption{\label{well.net.sand.fig} Interval picked to calculate the net sand in the well for Iris-1.}
\end{figure}

\begin{figure}
\noindent\includegraphics[width=20pc]{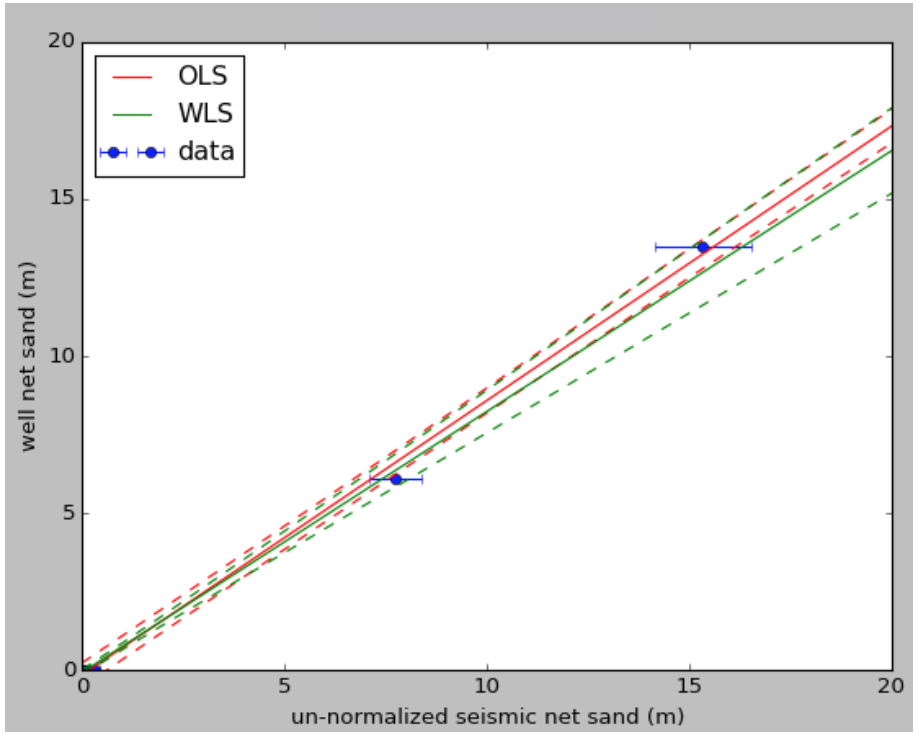}
\caption{\label{net.sand.reg.fig} Regression of the well net sand values versus the values estimated by the secant area after the first preliminary calibration for UP5/Iris. OLS = Ordinary Least Squares fit. WLS = Weighted Least Squares fit. Dotted lines show one standard deviation.}
\end{figure}

If one is in an exploration setting where the wells are from nearby fields, then one only has this initial calibration.  For the appraisal situation, where there are wells in the reservoir, an improved second calibration can be done.  For the second step of the calibration, a linear relationship was derived to correlate the net sand encountered in the well to the net sand derived from the secant area calculation after the first step of the calibration.  First, the percentage sand was integrated over the seismically resolvable interval for each of the wells, as shown in Fig. \ref{well.net.sand.fig}.  Note that any layers that are close to seismically interfering with each other are grouped together.  They can not be separated, because there would be a bias in their estimate if they would be analyzed separately then added together.  These values are then regressed versus the secant area net sand estimates (see Fig. \ref{net.sand.reg.fig}), and this linear correction is then applied to the net sand estimate.
%===============================%
%===============================%
%===============================%
%===============================%

What is meant by net sand is the thickness of the ``end-member'' sand.  It is assumed that the sand is a laminated mixture of an end-member sand and an end-member shale.  This view is supported by both the core photographs and conventional Thomas-Stieber crossplots.  To calculate volumetrics, this value of the net sand should be multiplied by the sand end-member porosity and the sand end-member gas saturation.  This is emphasized because porosity and saturation values used in other analyses are net-to-gross (i.e., sand fraction) averaged porosities that are significantly less than the end member porosity and saturation used in this analysis.

An ensemble of net sand maps was formed by taking the secant amplitude extraction for each of the 3SI inversions, and applying the two calibrations to each of those maps.  The net sand values were then sorted for each grid location.  This ensemble incorporates only the uncertainty in the seismic amplitudes.  It does not include the calibration uncertainty.  It also assumes an infinite spacial correlation, therefore it will be an upper bound on the integrated volumetric uncertainty.  The effects of lateral correlation will be addressed in Sec. \ref{lateral.correlation}.

\subsection{Building in lateral correlation}
\label{lateral.correlation}

It is important to build in the lateral correlation into the net sand ensemble before integrating the maps to obtain volumes.  The reason is that the standard deviation of the volumes will be reduced by a factor $R / \sqrt{A_m}$ due to the lateral correlation, and be further reduced by a factor of $\sqrt{1 - N R^2 / A_m}$ due to the well control and lateral correlation; where $R$ is the variogram range, $A_m$ is the area of the map, and $N$ is the number of wells. 

\begin{figure}
\noindent\includegraphics[width=20pc]{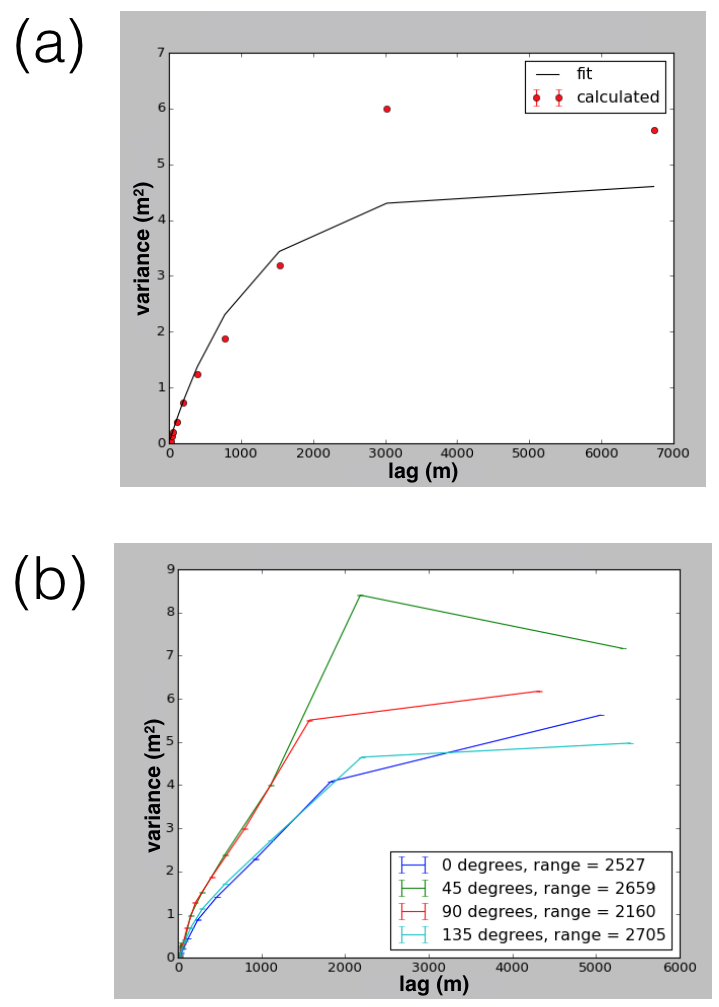}
\caption{\label{variogram.fig} Semivariogram analysis for UP5/Iris: (a) the fit variogram with anisotropy, (b) semivariogram analysis for the anisotropy.}
\end{figure}

Lateral correlation and well values are built into the ensemble of net sand maps using the method of \citet{gunning.et.al.07}.  Input to this method is a most likely net sand map, a standard deviation map, the variogram, along with the net sand encountered at each well location.  Output from the analysis are updated most likely and standard deviation maps, along an ensemble of correlated maps honoring the well control.  The anisotropic variogram is estimated by standard semivariogram analysis (on the most likely net sand map) as shown in Fig. \ref{variogram.fig}.  The well control comes from the well net sand points calculated from Fig. \ref{well.net.sand.fig}.  The input most likely net sand map is just the most likely doubly calibrated net sand map using the most likely wavelet to invert the unperturbed seismic data.  The standard deviation map is calculated two different ways.  The first is a standard deviation map of the ensemble of maps calculated in Sec. \ref{calibration}.  This is characteristic of the error in the seismic data and wavelet.  The second is a standard deviation map based on the error of the calibration shown in Fig. \ref{net.sand.reg.fig}, truncated so that the range does not include negative values of net sand.  This will be characteristic of the calibration error.
%===============================%
%===============================%

The volumetric distribution is calculated by taking a random map from the ensemble of correlated net sand maps using the seismic standard deviation, then integrating it over an area given by one of three contours chosen at random according to a probability of 20\% for the contour of smallest and largest area, and 60\% for the remaining contour.  This makes the contours the P10, P50, and P90 contours.  Porosity, gas saturation, gas expansion factor, rock volume norm (determined by the distribution of net sand volume from the ensemble of correlated net sand maps using the calibration standard deviation), and a non-pay discount factor (used to compensate for low saturation gas that is non-pay, but still produces an amplitude response) are chosen from suitably truncated normal distributions.  A Gas Initially In Place (GIIP) volume is calculated from these values.  An accounting of how much of this volume comes from each of the field areas is also done.  This process is repeated as many times as necessary to form a statistically significant ensemble of GIIP.  

A sensitivity analysis is done on the total GIIP to each of these factors.  Again, this sensitivity analysis is very valuable when it comes to making decisions about whether to drill additional wells, and where the wells should be located.  Specific examples of the Value Of Information (VOI) decisions based on this sensitivity analysis will be discussed in Sec. \ref{results}.

\section{Results}
\label{results}

The integrated analysis presented in Sec. \ref{theory} is applied to the Cassra/Iris field offshore Trinidad.  This is a gas field with six wells that were provided for this study -- two targeting the UP5 interval, and four that targeted primarily the M0 interval.  The quality of the well logs varied depending on their age.  Many of the wells had shear logs, but unfortunately most of the shear logs failed in these unconsolidated sections due to fundamental flaws in the shear logging tool design.  A well processed 3D seismic survey which had Bandwidth Extension \citep{smith.09} applied to it was available.  Only the full stack was used in this analysis.  It can be seen in Fig. \ref{stoch.rfc.fig} that there would be little value in a multiple stack AVO analysis for volumetric analysis.  This was further quantified by examining the confusion matrix for only the full stack and comparing it to confusion matrix for using the near and the far stack in Sec. \ref{net.pay.estimation}.  There was little change in this matrix when the near and far stacks were used with respect to discriminating gas from either brine or shale.

\begin{figure}
\noindent\includegraphics[width=20pc]{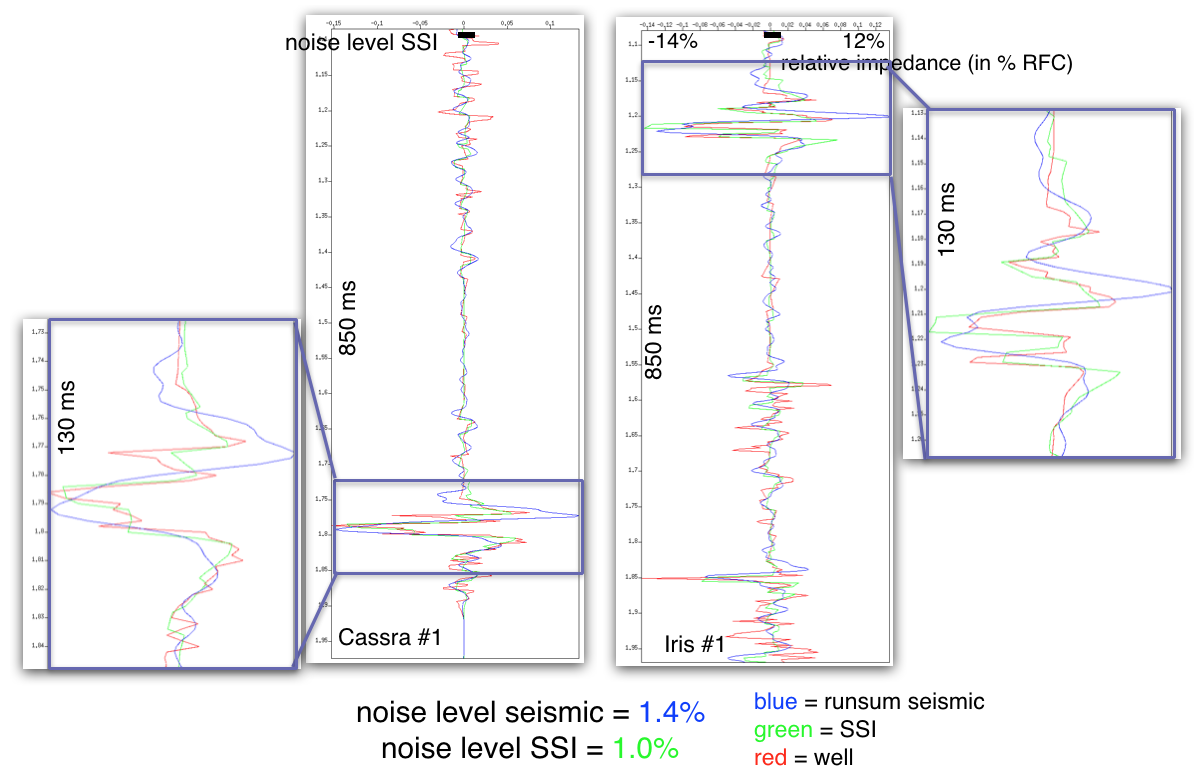}
\caption{\label{well.tie.fig} Comparison of the well acoustic impedance (red) to the seismic runsum (blue) and the sparse spike inversion impedance (green) for the Cassra-1 and Iris-1 wells.}
\end{figure}

\begin{figure}
\noindent\includegraphics[width=20pc]{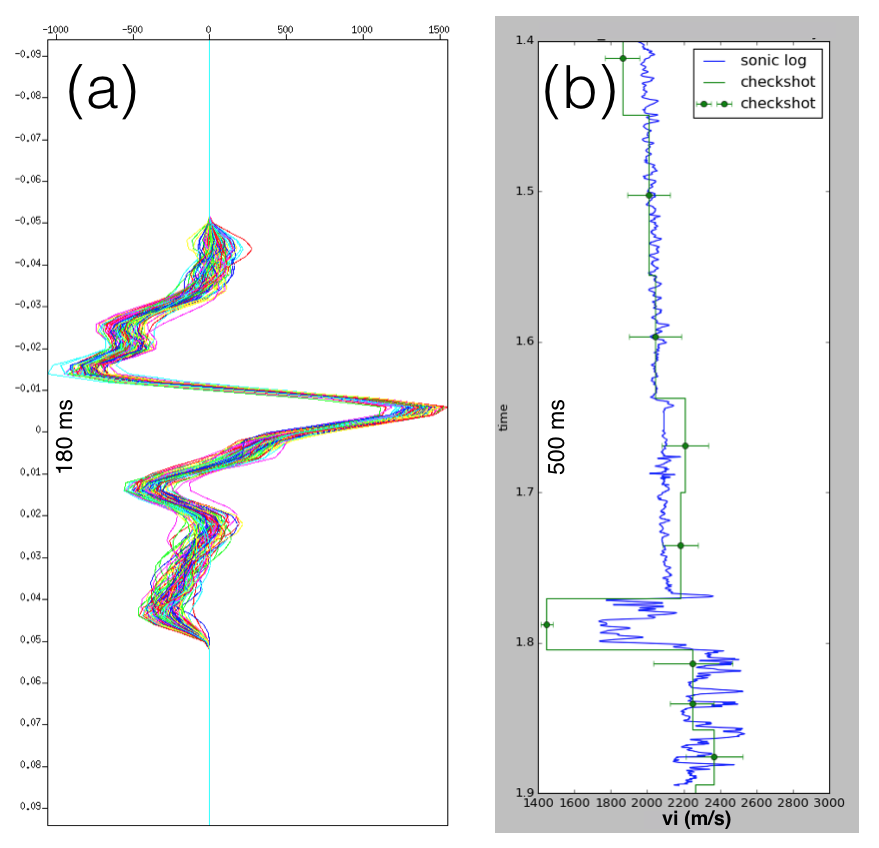}
\caption{\label{wavelet.fig} (a) Ensemble of wavelets demonstrating the uncertainty in the wavelet. (b) Time-to-depth mapping with uncertainty shown as interval velocity compared to the sonic well log for Cassra-1.}
\end{figure}

\begin{figure}
\noindent\includegraphics[width=20pc]{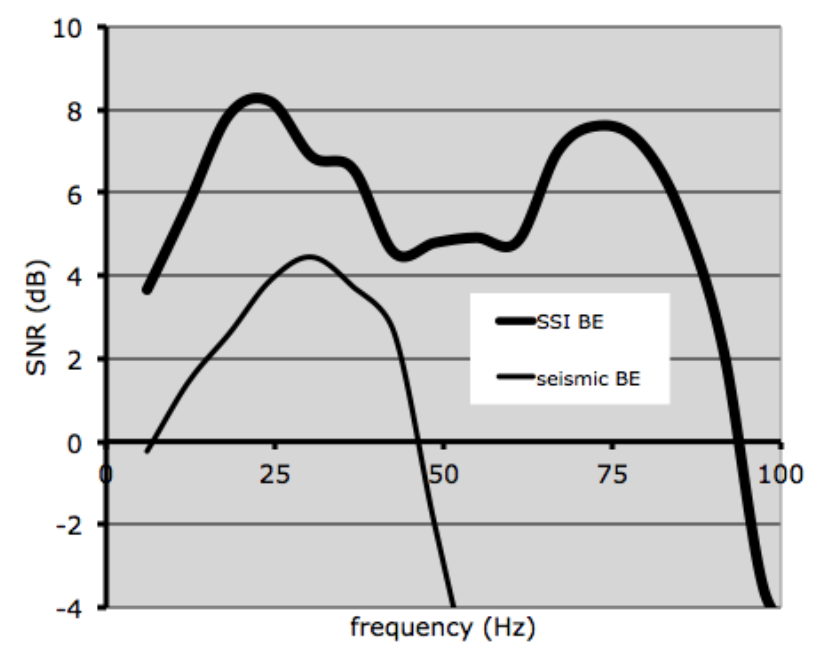}
\caption{\label{snr.fig} Spectral signal-to-noise-ratio (SNR) for the original Bandwidth Enhanced data, compared to the result of the sparse spike inversion.}
\end{figure}

\begin{figure}
\noindent\includegraphics[width=20pc]{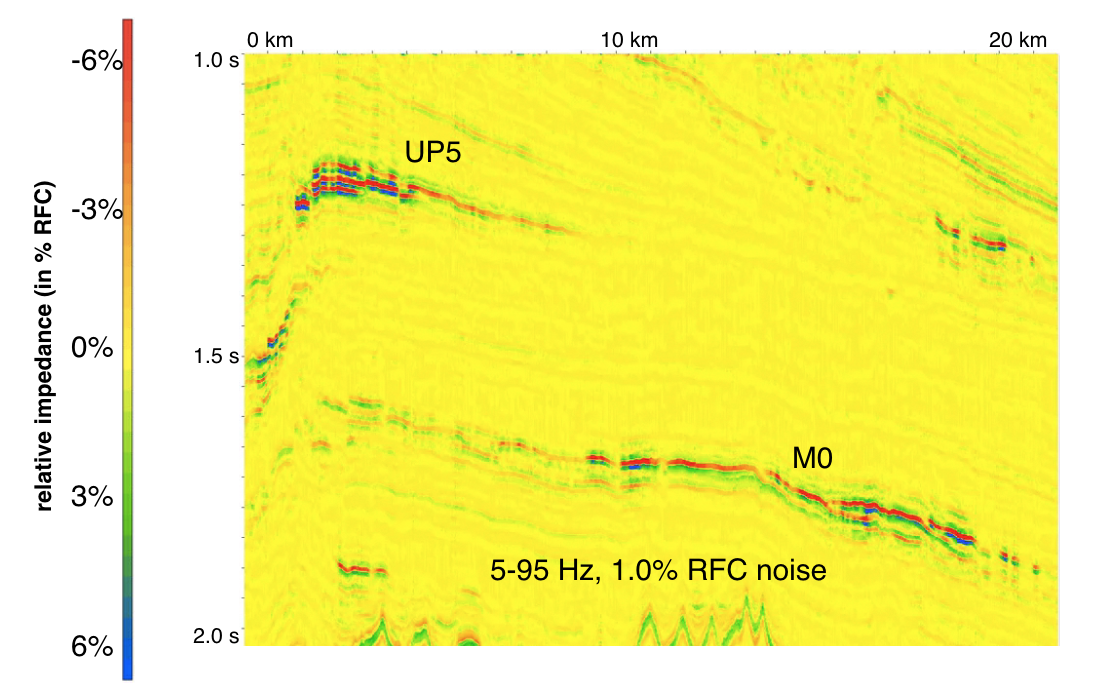}
\caption{\label{ssi.fig} Representative cross section of the sparse spike inversion impedance volume.}
\end{figure}

First, the two wells with the best logs, including a shear sonic (Cassra-1 and Iris-1) were used to do a wavelet derivation as described in Sec. \ref{wavelet.derivation}, followed by the stochastic sparse spike inversion described in Sec. \ref{stochastic.ssi} that generated 100 realizations.  The results are shown in Figs. \ref{well.tie.fig}, \ref{wavelet.fig}, \ref{snr.fig}, and \ref{ssi.fig}.  There is very good agreement between the well log and the result of the sparse spike inversion shown in Fig. \ref{well.tie.fig}.  This agreement is significantly better than the original seismic runsum data.  The noise level was improved from $1.4\%$ to $1.0\%$ in percent reflectivity units.  This should be compared to the $8.0\%$ to $12.0\%$ reflectivity of the the main target reflectors.  This is very good data.  This improvement is further quantified by Fig. \ref{snr.fig} which shows the spectral signal-to-noise (SNR) before and after the SSI.  Note that the useful bandwidth was increased from 10-45 Hz to 5-95 Hz by the SSI.  It should be noted that this large improvement would not have been possible if it was not for the Bandwidth Extension.  The Bandwidth Extension needed the SSI, in order to reach its full potential, because of the low frequency and ringy (i.e., many side lobes) wavelet that it leaves on the data (see Fig. \ref{wavelet.fig}).  The SSI removes this wavelet from the data.  Figure \ref{wavelet.fig} also shows the time-to-depth mapping with uncertainty that is output from the wavelet derivation.  Note the correspondence to the sonic log, even though there is no explicit constraint to it.  Finally a key cross section through the SSI is shown in Fig. \ref{ssi.fig}.  This shows very fine fault and stratigraphic detail.  Of particular note is the classic off-lapping clinoforms of the M0 deltaic sequence.
%===============================%
%===============================%
%===============================%
%===============================%
%===============================%
%===============================%
%===============================%
%===============================%

\begin{figure}
\noindent\includegraphics[width=20pc]{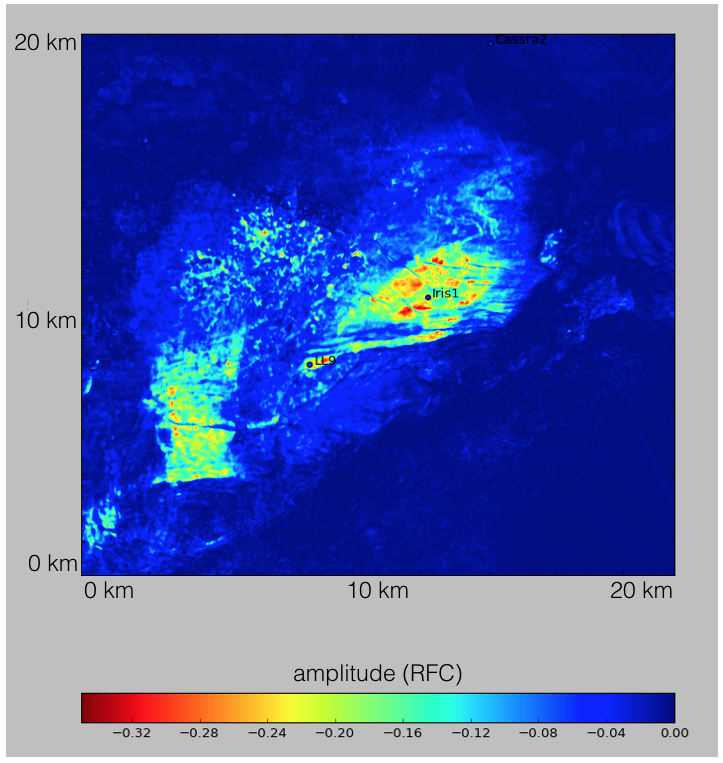}
\caption{\label{rfc.fig} Secant amplitude extraction from the UP5.  Units are reflection coefficient.}
\end{figure}

\begin{figure}
\noindent\includegraphics[width=20pc]{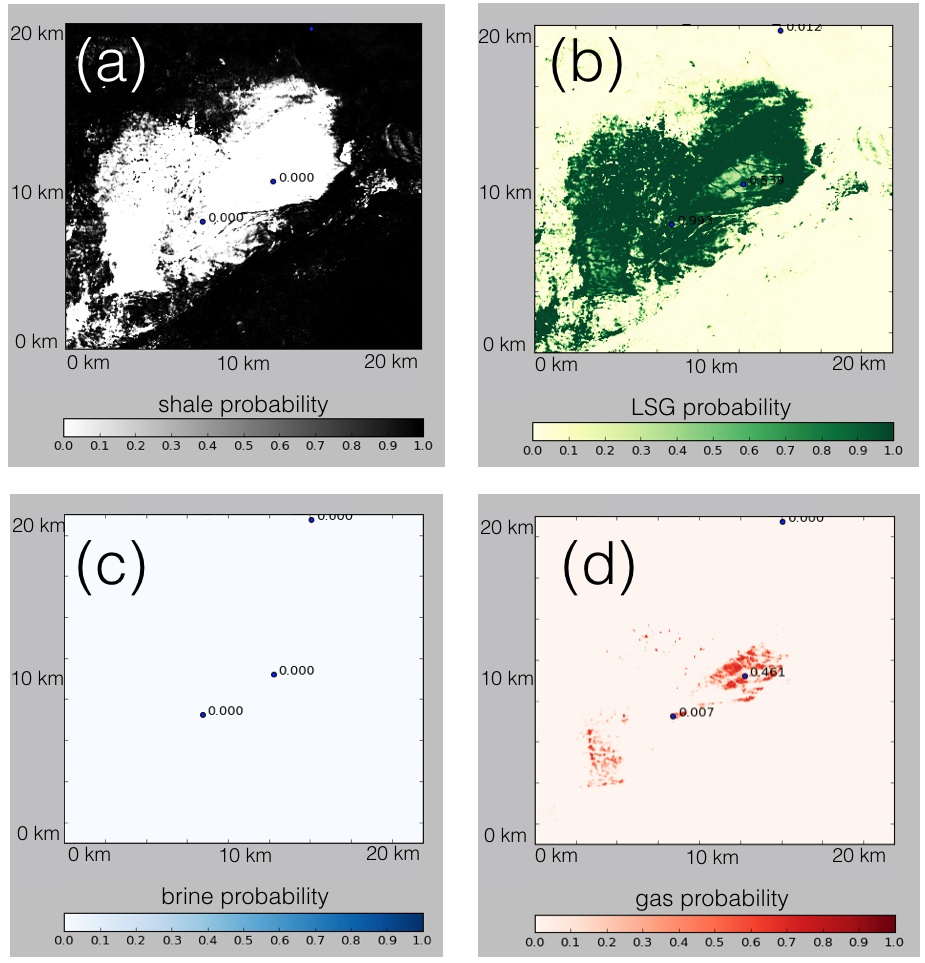}
\caption{\label{fluid.prob.fig} Fluid probability maps for the UP5:  (a) shale, (b) low saturation gas, (c) brine, (d) gas.  Note that the probability of low saturation gas is overestimated and the probability gas is underestimated, especially in the dim zone shown in Fig. \ref{dim.zone.fig}, due to the effects described in Sec. \ref{calibration}.}
\end{figure}

Stratigraphic interval specific work was done.  This paper presents the results for the UP5 turbidite sequence of the Iris reservoir.  As described in  Sec. \ref{theory}, a Stochastic Sparse Spike Inversion (3SI) was done based on the wavelet derivation, then an ensemble of secant amplitude and area maps were extracted from the ensemble of 3SI acoustic impedance volumes.  The most likely, calibrated amplitude map is shown in Fig. \ref{rfc.fig}.  This was used along with the conditional probabilities, shown in Fig. \ref{stoch.rfc.fig}, to generate the fluid probability maps, shown in Fig. \ref{fluid.prob.fig}.  

It should be noted that the sand needs to be of tuning thickness or greater.  Interpretation of these probabilities should recognize that sand below tuning or with a net-to-gross less than what was modeled will have secant amplitudes less than what would be modeled for that fluid.
%===============================%
%===============================%
%===============================%
%===============================%

Some comment should be made on the rock physics relationships that were used in the first part of the net sand calibration.  Because many of the sands and shales picked were not strict end members, the fit of the average was not representative of the end member (see Fig. \ref{rock.phys.fit.fig}).  To correct for this, the edge of the cluster of points was estimated and the orientation of the cluster determined.  The trend that was used represented the edge of the cluster, had the orientation of the cluster of points, and had a reduced standard deviation.  For the cases where there were not enough points to determine the orientation of the cluster, the orientation of global reference trends were used.

\begin{figure}
\noindent\includegraphics[width=20pc]{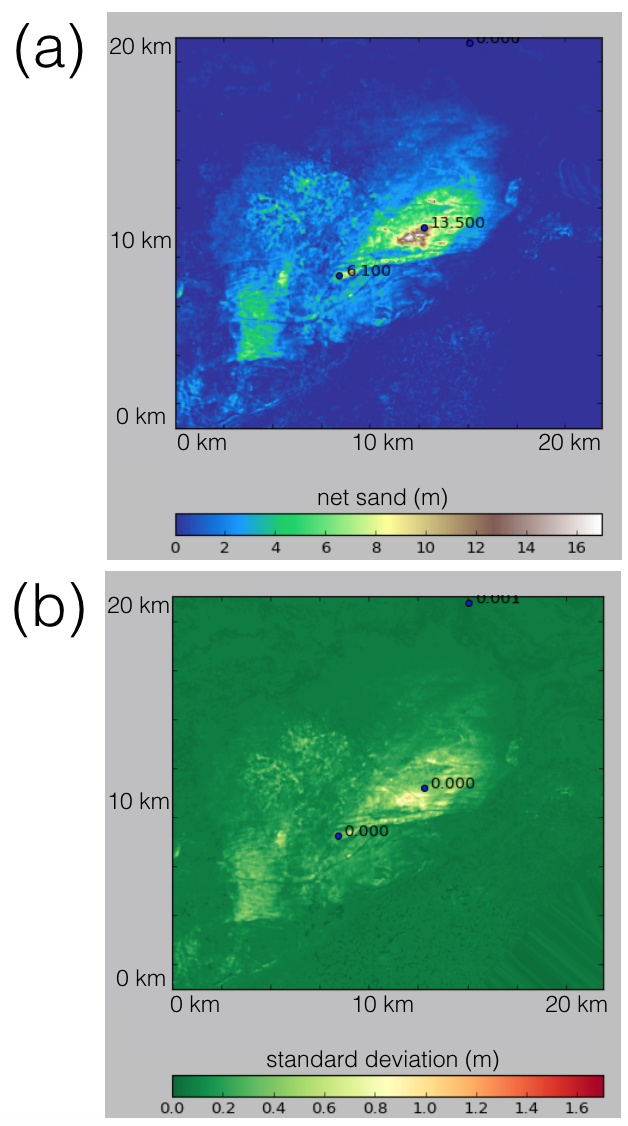}
\caption{\label{net.sand.map.fig} Calibrated net sand map for the UP5 with lateral correlation, tied to the well value:  (a) mean, (b) standard deviation.}
\end{figure}

\begin{figure}
\noindent\includegraphics[width=20pc]{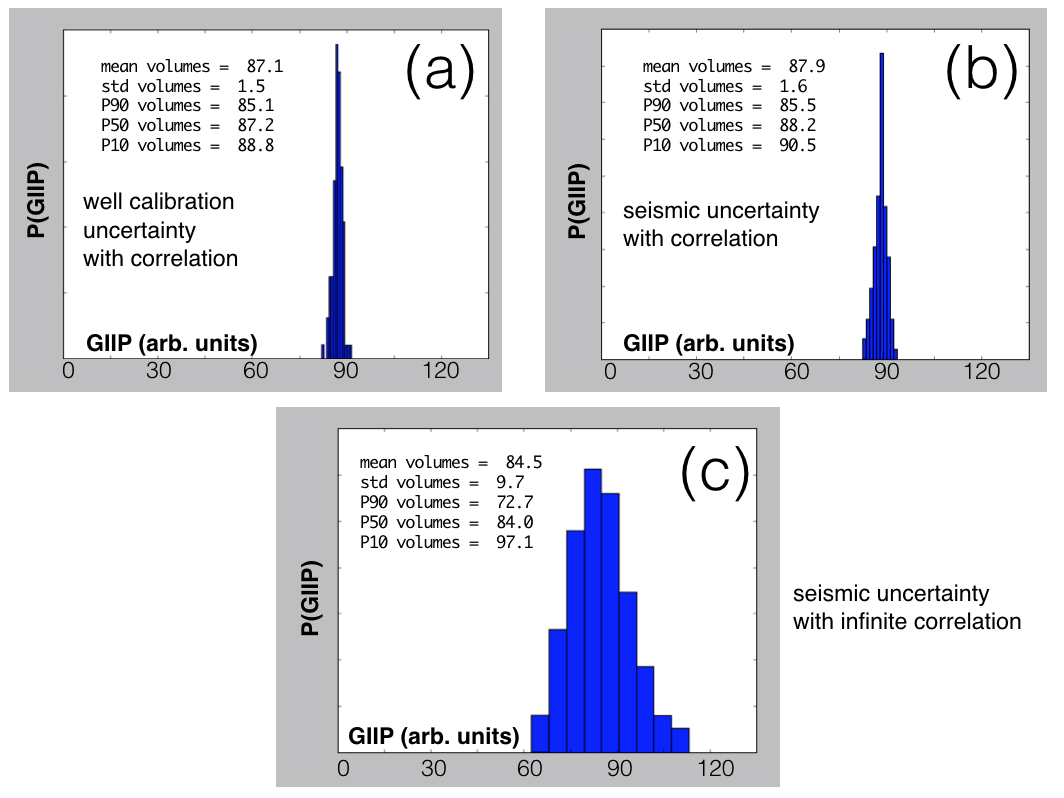}
\caption{\label{vol.fig} Preliminary volumetric distribution for UP5/Iris:  (a) well calibration uncertainty laterally correlated ensemble, (b) 3SI ensemble with lateral correlation, (c) raw 3SI ensemble with infinite lateral correlation.}
\end{figure}

Then, the secant amplitude ensemble was two step calibrated, laterally correlated, and tied to the wells to give the mean and standard deviation net sand maps shown in Fig. \ref{net.sand.map.fig}.  Preliminary evaluation of the volumetric distributions using the most likely contour and properties are shown in Fig. \ref{vol.fig} for three different map ensembles (infinite correlation raw 3SI ensemble, well calibration uncertainty laterally correlated ensemble, and the 3SI ensemble with lateral correlation).  The exponential variogram shown in Fig. \ref{variogram.fig} was used with a sill of $4.6 \, \text{m}^2$ and an anisotropic range of 3.4 x 2.2 km, 20 degrees north of east.  The range is the correlation length of the geologic process and is characteristic of the geologic body size.  An anisotropic range is simply a statement that the body shape is oblong.
%===============================%
%===============================%
%===============================%
%===============================%

\begin{figure}
\noindent\includegraphics[width=20pc]{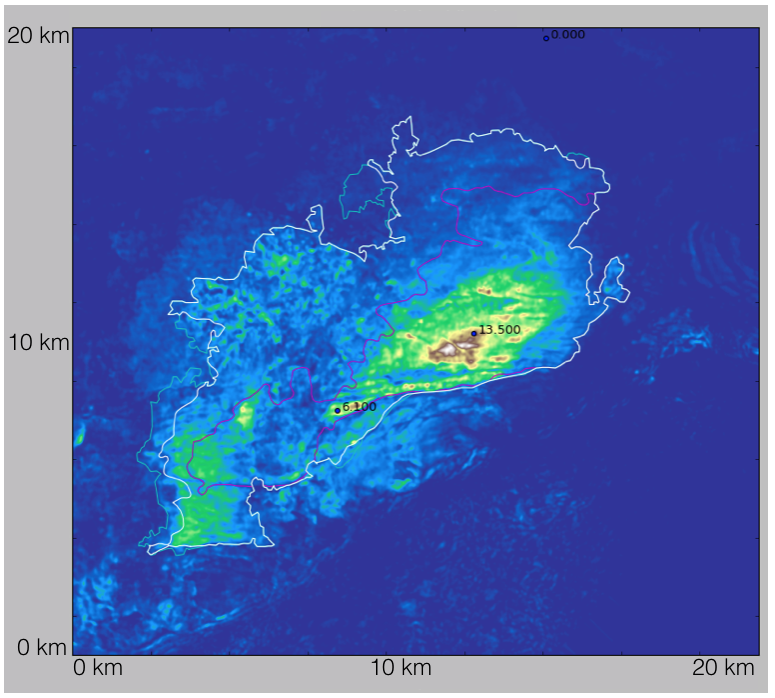}
\caption{\label{contours.fig} Contours used for volumetric integration for UP5/Iris (P90=magenta, P50=white, P10=cyan).}
\end{figure}

The three contours (P10, P50, and P90), shown in Fig. \ref{contours.fig} were used along with Gaussian distributions for porosity ($0.33 \pm 0.02$) , gas saturation ($0.85 \pm 0.04$), gas expansion factor ($115.0 \pm 1.0$), rock volume calibration ($1.00 \pm 0.04$), and non pay discount ($1.00 \pm 0.02$).  The ranges for porosity and gas saturation were determined by an analysis of the well logs for the end-member sands.  The range of the gas expansion factor was calculated using the equation of state and the estimated ranges for temperature and pressure.  The rock volume calibration was estimated by using Fig. \ref{vol.fig}a (the fractional width of the distribution).  The non-pay discount accounted for known low saturation gas which would contribute to the seismically estimated net pay.
%===============================%
%===============================%

\begin{figure}
\noindent\includegraphics[width=20pc]{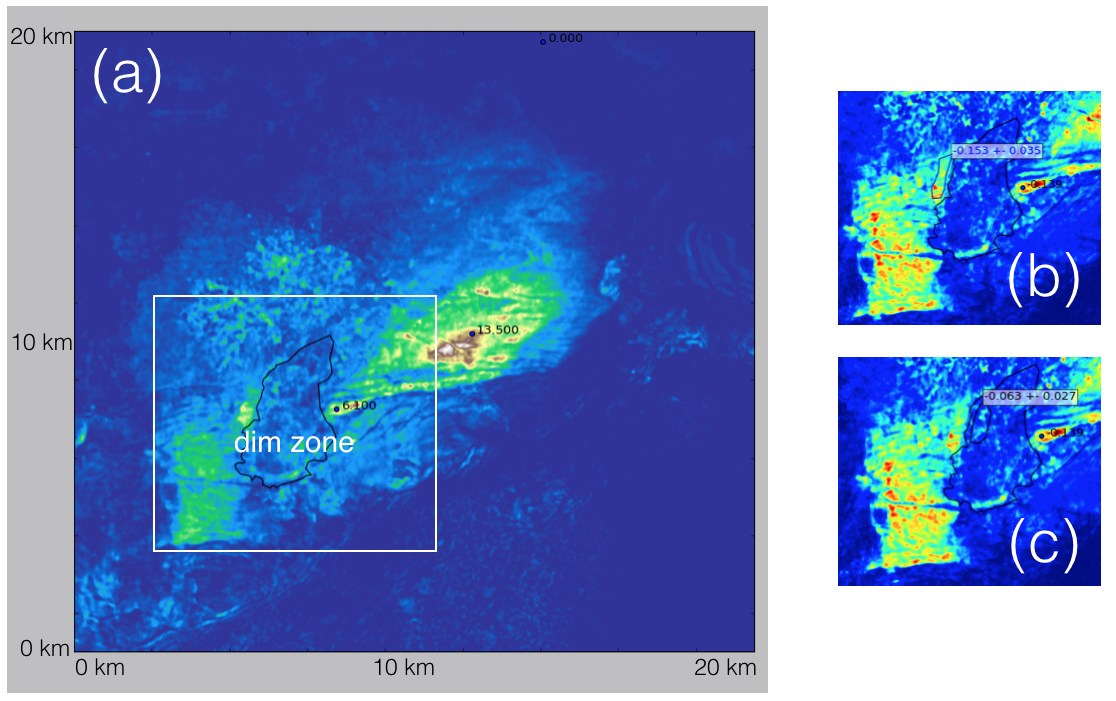}
\caption{\label{dim.zone.fig} Contour used to define the dim zone in the UP5:  (a) extent of the shadowed area on the net sand map, (b) area of un-shadowed amplitude shown on the reflection coefficient map, (c) area of shadowed amplitude shown on the reflection coefficient map.  Amplitude ratio is $0.153/0.063=2.4$.}
\end{figure}

Part of the reservoir at this stratigraphic interval is in the shadow of some shallow gas.  This dim zone, shown in Fig. \ref{dim.zone.fig}, is compensated for by identifying an area that is not shadowed and a stratigraphically similar area that is shadowed.  The difference in the mean amplitudes of these areas is calculated and the volume of the area of the dim zone is multiplied by this factor.  
%===============================%
%===============================%

\begin{figure}
\noindent\includegraphics[width=20pc]{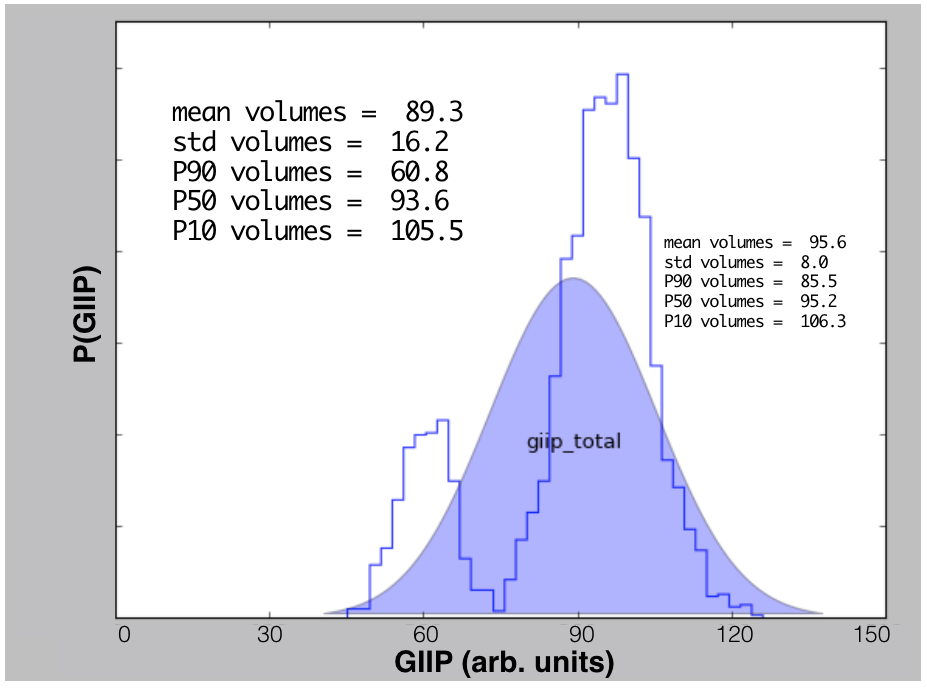}
\caption{\label{vol.dist.fig} Integrated volumetric distributions for UP5/Iris.  The dark, blocky, line is the blocked distribution function.  The smooth, transparently shaded curve is a fit Gaussian distribution.}
\end{figure}

The resulting volumetric distributions (with 3000 realizations) that combine all these factors are shown in Fig. \ref{vol.dist.fig}.  Note that this distribution is multimodal.  The difference between these two modes is the area between the P50 and the P90 contours. The results of these volumetrics are shown in Table \ref{iris.vol.table}.
%===============================%
%===============================%

%===============================%
\begin{table}
\centering
\begin{tabular}{|r||cc|ccc|}
\hline
unit & \; mean \; & \; std dev \;  & \; P90 \; & \; P50 \; & \; P10 \; \\
\hline \hline
$\mathbf{Iris (total)}$ & 89.2 & 16.2 & 60.8 & 93.6 & 105.5\\
Iris (uncorrected) & 81.7 & 15.0 & 55.8 & 85.6 & 97.2\\
dim zone & 7.5 & 1.4 & 5.1 & 7.9 & 9.1\\
\hline
Iris (P50 contour) & 95.6 & 8.0 & 85.5 & 95.2 & 106.3\\
\hline
\end{tabular}
\caption{Volumetric distributions of the UP5 reservoir interval.  Units are arbitrary.  Dim zone volume is the additional volume added by the correction. }
\label{iris.vol.table}
\end{table}
%===============================%

\begin{figure}
\noindent\includegraphics[width=20pc]{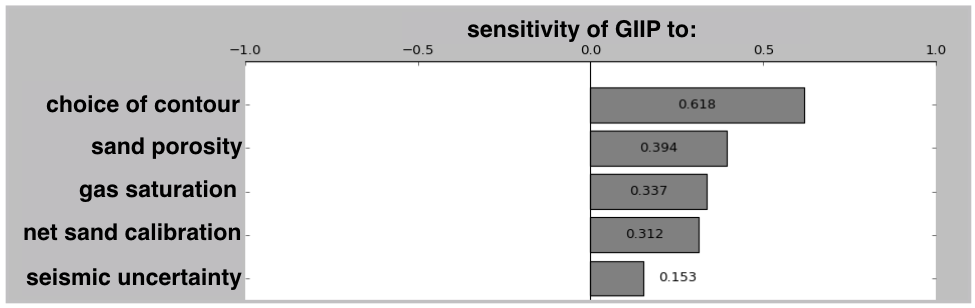}
\caption{\label{vol.sensitivity.fig} Sensitivity analysis, shown as tornado chart, of the UP5/Iris GIIP to the independent stochastic variables.}
\end{figure}

The sensitivity of the Iris volumes is shown in Fig. \ref{vol.sensitivity.fig}.  This is an important figure that warrants some further discussion.  The dominate input uncertainties are the choice of contour, sand porosity, gas saturation, and the net sand calibration (in that order).  A well into the area between the P50 and P90 contour that was never penetrated would be the best way to reduce the dominate uncertainty in the choice of contour.  Next a more careful petrophysical analysis could be done to determine the range of the porosity and gas saturation of the end member sand.  Finally, an additional well could be drilled in the thickest part of the reservoir to improve the calibration of the net sand maps.
%===============================%
%===============================%

\section{Conclusions}
\label{conclusions}

A state-of-the-art stochastic wavelet derivation was applied, giving the seismic noise level, time-to-depth mappings, and the wavelet -- all with uncertainty.  This was used to do a novel stochastic sparse spike inversion.  An ensemble of secant area maps were extracted from the ensemble of sparse spike inversion impedance volumes.  The secant area maps were calibrated to net sand maps in a two step process.  Lateral correlation was added to this ensemble of net sand maps and they were tied to the well values of net sand.  Finally, uncertainty in the contour area, gas saturation, porosity, and gas expansion factor were added to give an integrated view of the volumetric uncertainty.  A sensitivity analysis was done to give insight into the value of information -- what data should be acquired and what wells should be drilled in order to maximize the return on capital investment.

%% End of body of article:

%%%%%%%%%%%%%%%%%%%%%%%%%%%%%%%%%%%%%%%%%%%%%%%%%%%%%%%%%%%%%%%%
%
%  ACKNOWLEDGMENTS

\begin{acknowledgments}

The authors would like to thank Centrica E\&P for supporting this work, supplying the Cassra/Iris Field data, and granting permission to publish.  The contributions of Petrotrin, as the joint venture partner, and the Ministry of Energy and Extractive Industries (MEEI) are recognized and appreciated.  We would also like to thank Geotrace Technologies for their financial support, and intellectual encouragement.

\end{acknowledgments}

%% ------------------------------------------------------------------------ %%
%%  REFERENCE LIST AND TEXT CITATIONS
%

%merlin.mbs aipnum4-1.bst 2010-07-25 4.21a (PWD, AO, DPC) hacked
%Control: key (0)
%Control: author (8) initials jnrlst
%Control: editor formatted (1) identically to author
%Control: production of article title (-1) disabled
%Control: page (0) single
%Control: year (1) truncated
%Control: production of eprint (0) enabled
%

\end{document}